# Freeform terahertz structures fabricated by multi-photon lithography and metal coating

Pascal Maier,[1,2] Alexander Kotz,[1] Joachim Hebeler,[3] Qiaoshuang Zhang,[4] Christian Benz,[1,2] Alexander Quint,[3] Marius Kretschmann,[3] Tobias Harter,[1] Sebastian Randel,[1] Uli Lemmer,[4] Wolfgang Freude,[1] Thomas Zwick,[3] and Christian Koos[1,2,5,*]

[1]*Institute of Photonics and Quantum Electronics (IPQ), Karlsruhe Institute of Technology (KIT), Engesserstr. 5, 76131 Karlsruhe, Germany*
[2]*Institute of Microstructure Technology (IMT), KIT, Hermann-von-Helmholtz-Platz 1, 76344 Eggenstein-Leopoldshafen, Germany*
[3]*Institute of Radio Frequency Engineering and Electronics (IHE), KIT, Engesserstr. 5, 76131 Karlsruhe, Germany*
[4]*Light Technology Institute (LTI), KIT, Engesserstr. 13, 76131 Karlsruhe, Germany*
[5]*Vanguard Photonics GmbH, Gablonzer Str. 10, 76185 Karlsruhe, Germany*
* christian.koos@kit.edu

**Abstract:** Direct-write multi-photon laser lithography (MPL) combines highest resolution on the nanoscale with essentially unlimited 3D design freedom. Over the previous years, the groundbreaking potential of this technique has been demonstrated in various application fields, including micromechanics, material sciences, microfluidics, life sciences as well as photonics, where *in-situ* printed optical coupling elements offer new perspectives for package-level system integration. However, millimeter-wave (mmW) and terahertz (THz) devices could not yet leverage the unique strengths of MPL, even though the underlying devices and structures could also greatly benefit from 3D freeform microfabrication. One of the key challenges in this context is the fact that functional mmW and THz structures require materials with high electrical conductivity and low dielectric losses, which are not amenable to structuring by multi-photon polymerization. In this work, we introduce and experimentally demonstrate a novel approach that allows to leverage MPL for fabricating high-performance mmW and THz structures with hitherto unachieved functionalities. Our concept exploits *in-situ* printed polymer templates that are selectively coated through highly directive metal deposition techniques in combination with precisely aligned 3D-printed shadowing structures. The resulting metal-coated freeform structures (MCFS) offer high surface quality in combination with low dielectric losses and conductivities comparable to bulk material values, while lending themselves to *in-situ* fabrication on planar mmW and THz circuits. We experimentally show the viability of our concept by demonstrating a series of functional THz structures such as ultra-broadband chip-chip interconnects, THz probe tips, and suspended THz antennas. We believe that our approach offers disruptive potential in the field of mmW and THz technology and may unlock an entirely new realm of laser-based 3D manufacturing.

## 1. Introduction

Functional terahertz (THz) structures crucially rely on precisely defined three-dimensional (3D) freeform geometries that combine highly conducting metal elements with low-loss dielectrics. Geometrical precision can be achieved by direct-write multi-photon laser lithography (MPL)[1,2], offering highest resolution on the sub-micrometer scale along with precise alignment of the fabricated structures with respect to existing circuitry on the underlying substrate. These advantages have been extensively exploited in the realm of photonic integration[3], where 3D-printed waveguides, so-called photonic wire bonds (PWB)[4–7], or facet-attached microlenses (FaML)[8,9] offer interesting perspectives for scalable fully automated assembly of hybrid multi-chip systems. Transferring these concepts to THz assemblies, however, has so far been hindered by the lack of microfabrication techniques that can complement 3D-printed polymeric base structures by precisely defined highly conductive metal elements[3,10–12]. Specifically, while two-photon-induced reduction of metal salts has been exploited to generate free-standing 3D metal structures[13,14] or to decorate the surfaces of 3D-printed polymer structures with local silver[15–17] or platinum[12] lines, the resulting surface quality and the conductivity of the metal parts are still insufficient for low-loss millimeter-wave (mmW) or THz devices. Similar chemical reaction mechanisms can be used for fabricating gold-containing nanocomposite structures with 3D freeform shapes[18]. However, the achievable electrical conductivity is limited by the metal loading, which is generally dictated by the solubility of the metal ions[11] in the respective photoresist. Moreover, when the incident laser beam irradiates already fabricated metal structures during printing, thermal effects and localized surface plasmon resonances can lead to aggregation and crystallization of metal nanoparticles, rendering these processes hard to control. It should also be noted that all the aforementioned techniques have so far been limited to a rather small selection of metals such as silver[13,15–17,19–23], gold[14,18,24–27], palladium[22,28,29], and platinum[12,23,28]. On the other hand, a combination of physical vapor deposition (PVD) techniques and subsequent electroplating of the PVD seed layer has been used for globally covering 3D-printed polymer templates with highly conductive metal coatings[30–32]. However, this concept has mainly been limited to bulky stand-alone hollow-core waveguides[30] or horn antennas[31,32] and often requires additional mechanical assembly steps[31] to produce functional mmW or THz elements. Localized metal coatings can be fabricated on 3D-printed polymer templates by means of chemical surface functionalization followed by electro-less plating[15,33,34], but this approach often suffers from poor coating homogeneity and rather low conductivity of the fabricated metal structures[10] and is hence not well suited for high-performance mmW or THz elements.

In this paper, we introduce and experimentally demonstrate a novel concept for fabricating precisely defined mmW and THz structures that combine low-loss dielectrics and highly conductive metal elements in a well-defined 3D freeform geometry. Our concept exploits *in-situ* printed polymer support structures that are selectively coated through highly directive metal deposition techniques in combination with precisely aligned 3D-printed shadowing structures. The resulting metal-coated freeform structures (MCFS) offer high surface quality in combination with conductivities comparable to bulk material values and do not require any manual assembly steps. We demonstrate the viability of the concept in a series of experiments. In the first set of experiments, we use our approach to demonstrate THz interconnects that bridge the gap between transmission lines located on different substrates and that offer 3 dB-bandwidths exceeding 0.33 THz. In a second set of experiments, we show that the vast design freedom offered by our fabrication technique can be leveraged for cost-effective THz probe tips that allow highly repeatable contacting over many probing cycles with 3 dB-bandwidths exceeding 0.19 THz and with 6 dB-bandwidths far beyond the 0.33 THz range of our measurement system. A third set of experiments is finally dedicated to THz antennas, which are suspended from the underlying high-index substrate for better radiation efficiency. At a frequency of 0.27 THz, a maximum realized gain of 5.5 dBi in the direction perpendicular to the substrate is measured. The presented proof-of-principle experiments demonstrate the vast potential offered by the proposed fabrication technique. Our concept offers unprecedented design freedom and is widely applicable to a rich variety of use cases, thereby paving a path towards advanced mmW and THz systems in communications[35], sensing[36], or ultra-broadband signal processing[37].

## 2. Concept

An exemplary THz system that exploits 3D-printed functional elements is illustrated in Fig. 1. The example depicted in Fig. 1a combines an optoelectronic signal source, relying on a photonic integrated circuit (PIC) with a balanced pair of high-speed photodetectors, with a subsequent THz amplifier, based on a monolithic microwave integrated circuit (MMIC), and a THz antenna, which is suspended from the surface of the underlying substrate for efficient emission to the surface-normal direction. The balanced photodetectors are co-integrated with other functional photonic structures such as multi-mode-interference (MMI) couplers that are fed with an optical signal and an optical local-oscillator (LO) tone supplied by an array of single-mode fibers (SMF). The THz chip-chip interconnect between the PIC and MMIC as well as the suspended THz antenna are implemented as metal-coated freeform structures (MCFS). Each MCFS consists of a 3D-printed polymeric support that is locally coated with metal layers offering high bulk conductivity along with mmW-grade surface quality. The same concept can be used to implement THz probes with unprecedented shape fidelity that lend themselves to testing of mmW and THz integrated circuits, see upper part of Fig. 1a. The polymeric support structures of the various MCFS are fabricated using high-resolution multi-photon lithography (MPL) and can thus be efficiently combined with 3D-printed optical connections such as photonic wire bonds (PWB). The viability of the concept shown in Fig. 1a is demonstrated in a series of proof-of-concept experiments – a selection of fabricated MCFS-based functional THz structures is shown in Fig. 1b, c and d. More specifically, Fig. 1b shows a THz interconnect (TIC) bridging the gap between coplanar waveguide (CPW) transmission lines on different THz substrates, Fig. 1c shows a THz probe (TP) that is turned upside down for better visibility, and Fig. 1d shows a THz antenna (TA) that is suspended from the underlying high-index substrate for increased radiation efficiency.

The fabrication of the MCFS relies on a dedicated multi-step process that is illustrated in Fig. 2. The process steps are shown exemplarily for the fabrication of a TIC, see Fig. 1b. In a first step, two substrates with prefabricated planar THz structures such as CPW are coarsely placed to face each other with a gap in-between, see Fig. 2a. As a basis for later directive metal coating, illustrated in Fig. 2d, support and shadowing structures are 3D-printed using *in-situ* direct-write MPL, see Fig. 2b. The structures are designed to support a metal layer which connects smoothly to the CPW on each substrate and comprise isolation trenches with undercut sidewalls to separate the deposited ground-signal-ground (GSG) metal strips, see Insets in Fig. 2b and d. For protecting the planar THz circuits during global metal deposition, the associated areas of the substrates are temporarily covered by a poly(methyl methacrylate) (PMMA) film using inkjet printing[38], Fig. 2c. At the transitions between the PMMA-covered planar substrates and the 3D freeform polymeric support, additional 3D-printed shadowing structures are used to prevent short circuits. These shadowing structures, which are shaded in blue in Fig. 2b – e, take the form of multiple arm-like roofs which locally prevent metal deposition, thereby defining isolating regions that separate the metal strips of the GSG transmission line and that seamlessly transition to the isolation trenches of the adjacent 3D support structure. Inset ⓣ in Fig. 2c provides a top-view on one side of the support structure with the corresponding shadowing structures above – the isolation regions protected from metal deposition are shaded in blue. The shadowing structures are built upon a box-like hollow base, which acts as a flow-stop preventing inkjet-printed PMMA from wetting the entire support structure. In a next step, a highly directed physical vapor deposition (PVD) process is used to deposit metal along a surface-normal direction, see Fig. 2d, thereby forming highly conductive layers on all surfaces with direct line of sight to the evaporation source. After metal deposition, the sacrificial PMMA layer is dissolved, thereby lifting off the unwanted metal areas, see Fig. 2e. As a last step, the 3D-printed shadowing structures can be removed mechanically, leaving the final MCFS as shown in Fig. 2f.

More details on the processes and the design considerations of the structures can be found in the Methods and in Supplementary Section 2. Specifically, the width of the signal and ground conductors $w_S$ and $w_G$ as well as the associated distance $d_{SG}$ of the TIC-based CPW were chosen to provide a desired line impedance of, e.g., 50 Ω along the entire TIC structure. To this end, the dimensions of the 3D-printed freeform support structure and the associated isolation trenches can be continuously varied along the propagation path of the signal, e.g., by continuously adapting the width of the gaps $w_{gap}$, which defines the distance $\approx d_{SG}$ between the signal and ground conductors at the respective height $h$, see Inset in Fig. 2b and d. The metal residues which are deposited at the ground of the isolation trenches, see Inset in Fig. 2d, do not influence the characteristics of the MCFS-based TIC if the undercuts are designed to offer sufficient depth $d_{it}$.

To verify the high conductivity and surface quality of the deposited metal layers, we analyzed various fabricated MCFS. The surface roughness was measured using a white-light interferometer, revealing a root-mean-square (RMS) surface roughness of $R_{q,MCFS} = (13...14)\,\text{nm}$, slightly larger than the roughness of the underlying 3D-printed support of $R_{q,support} = (9...10)\,\text{nm}$, see Supplementary Section 3 for further information. The conductivity of the metal films was extracted from four-wire measurements[39] of metal strips which were separately fabricated on an oxidized silicon wafer using identical evaporation processes and metal layer stacks, see Methods section below. Assuming a homogeneous metal layer, we extract an effective conductivity of $\sigma_{MCFS} = (3.29 \pm 0.12) \times 10^7\,\text{S/m}$. This corresponds to $(57 \pm 2)\%$ of the bulk material value of copper and is consistent with previously reported values using similar deposition techniques[40]. The slight conductivity reduction compared to bulk material is a known effect related to the grain boundaries of the deposited metal layers[40], which might be mitigated by further process optimization.

### 3. Experimental verification and discussion

To verify the viability of our concept and to quantify the associated performance parameters, we performed a series of experiments which were geared towards demonstration of the building blocks shown in Fig. 1b, c and d: THz interconnects (TIC) bridging the gap between transmission lines on different substrates, THz probes (TP) that allow repeatable contacting of THz circuits, and THz antennas (TA) that are suspended from the underlying high-index substrate for increased radiation efficiency. These experiments are discussed in the following sections.

*THz interconnects (TIC)*

A micrograph of three manufactured TIC connecting three pairs of CPW on two separate alumina substrates is shown in Fig. 3. For all structures, the CPW endings are separated by $L_{\text{TIC}} \approx 0.5\,\text{mm}$ and seamlessly connect to the respective TIC, labeled I, II, and III in Fig. 3b. For characterization of the TIC, we first measure the compound *S*-parameters of the TIC and the associated feed lines, which are contacted using dedicated probes in Plane $1'$ and Plane $2'$, indicated as blue lines in Fig. 3a. The length of the feed lines between the probes and the TIC amounts to $L_{\text{feed}} = 4\,\text{mm}$. The measured compound *S*-parameters are exemplarily depicted for TIC II in Fig. 3c, where the blue curve refers to the transmission $S_{2'1',\text{dB}} = 10\log_{10}(|\underline{S}_{2'1'}|^2)$ and the green curve to the reflection $S_{1'1',\text{dB}} = 10\log_{10}(|\underline{S}_{1'1'}|^2)$. The measurements show excellent agreement with simulations (dashed black lines), see Methods for details. To remove the influence of the feed lines, the reference planes need to be moved to Plane 1 and Plane 2, indicated as red lines in Fig. 3a and b. To this end, we separately measure the *S*-parameters of a reference CPW with identical cross section and a length of $L_{\text{feed}}$, which was fabricated on the same substrate. The *S*-parameters of the reference CPW are then used to de-embed the scattering parameters of the TIC, see Supplementary Section 4. Figure 3d shows the associated transmission characteristics $S_{21,\text{dB}} = 10\log_{10}(|\underline{S}_{21}|^2)$ (red lines) for TIC I (upper), TIC II (center) and TIC III (lower), revealing 3 dB-bandwidths of 0.307 THz, >0.330 THz and 0.290 THz, respectively. We again performed simulations (dashed black lines), which agree very well with the measurements. Note that the measured *S*-parameters had to be acquired separately in different frequency ranges using dedicated signal sources, waveguides, and probes, thereby leaving a gap between 0.170 THz and 0.200 THz, where no adequate signal sources were available. Note further that the depicted results rely on a de-embedding procedure using the transfer-matrix (*T*-matrix) approach[41] for frequencies up to 0.170 THz, while, for frequencies beyond 0.200 THz, the de-embedded *S*-parameters were estimated using a more robust scalar correction that is less prone to error multiplication. More details on the measurement setup and the de-embedding procedures are given in Supplementary Section 4. To the best of our knowledge, the demonstrated 3 dB-bandwidths in excess of 0.33 THz represent a record for in-plane mmW chip-chip connections and can already well compete with advanced CPW-based flip-chip interfaces[42,43]. Competing approaches comprise aerosol-jet-printed conductive lines, deposited on dielectric ramps[44–47] or epoxy underfills[44,48], as well as lithographically defined self-aligning metal nodules, that protrude from the facets of the chips and that require subsequent reflow fusing or electroless plating processes[49]. However, these approaches have only been demonstrated to work up to 0.22 THz in few cases[47,49] and generally lack the precision and design flexibility offered by the presented MCFS concept.

*THz probes (TP)*

We further used the MCFS concept to fabricate broadband TP. The measurement setup for characterization of the TP is shown in Fig. 4a. For the experiment, identical twins of TP ($L_{\text{TP}} = 415\,\mu\text{m}$, $h = 180\,\mu\text{m}$) are fabricated on a sample chip at opposite sides of a CPW ($L_{\text{conn}} = 3.8\,\text{mm}$). The three tips (pitch 100 µm, contact pad size $20 \times 40\,\mu\text{m}^2$) protrude beyond the edges of the sample chip by a distance $d_{\text{edge}}$. A second chip with two short CPW feeds ($L_{\text{feed}} = 1.5\,\text{mm}$) was used for contacting, where the distance between the endings of the CPW are slightly smaller than $L_{\text{TP}} + L_{\text{conn}} + L_{\text{TP}}$. A vacuum chuck attached to a manual actuator moves the sample chip in the desired position above the contacting chip. With the sample chip in contact, *S*-parameters between Plane $1''$ and Plane $2''$ (brown lines in Fig. 4a) are measured. Using again dedicated de-embedding procedures, the reference planes are first moved to Plane $1'$ and Plane $2'$, corresponding to the blue lines in Fig. 4a, see Supplementary Section 5 and associated Fig. S6 for details. In a second step, we move the reference planes to Plane 1 and Plane 2 (red lines in Fig. 4a), thereby estimating the characteristics of a single TP – the associated transmission factor $S_{21,\text{dB}} = 10\log_{10}(|\underline{S}_{21}|^2)$ is shown in Fig. 4b. The 3 dB-bandwidth amounts to 0.192 THz, and the 6 dB-bandwidth is far beyond the 0.330 THz range of our measurement system. Measurement and simulation (dashed black line) are again in reasonable agreement. We attribute the slight deviations above 0.1 THz to the probes used to connect the contacting chip to our measurement equipment. Specifically, these probes are intended for wafer-level measurements with surface qualities superior to the comparatively rough gold strips of the CPW feeds on our test substrates. More details on the measurement setup and the de-embedding procedures are given in Supplementary Section 5.

To the best of our knowledge, the demonstrated 3 dB-bandwidths in excess of 0.19 THz represent a record for additively manufactured electrical probes. Previous demonstrations were limited to the (75–110) GHz frequency band, with insertion losses of the order of 1 dB per probe[50]. To further demonstrate the robustness of our probes and to verify the repeatability of the connection, we repeat our measurement $N = 50$ times, detaching and re-establishing the contact between the sample and the contacting chip after each repetition. From the resulting data, we extract the standard deviation of the magnitude of the complex-valued amplitude transmission factor $\underline{S}_{21}$ as a function of frequency, see Fig. 4c. We find a remarkably low maximum standard deviation of $\sigma(|\underline{S}_{21}|)_{\max} \approx 2 \times 10^{-3}$ near 0.32 THz, corresponding to −55 dB. These results are already on par with conventional commercially available probes fabricated via conventional micromachining techniques where, e.g., standard deviations $\sigma(|\underline{S}_{21}|)$ in the range (−60…−43) dB have been demonstrated for 2 mm-long CPW in the 0.3 THz band[51].

*THz antennas (TA)*

As a last implementation example, we explore TA. A schematic of a manufactured TA operating at a frequency centered around 0.28 THz is shown in Fig. 5a. Connecting to the CPW feed ($L_{\text{feed}} = 1.1\,\text{mm}$), the 3D-printed support structure provides a mechanical base for guiding the signal and ground conductors of the CPW away from the substrate to form an elevated slot antenna for improved radiation into the upper half-space. The antenna comprises two slots, each having a width of $w_{\text{slot}} = 60\,\mu\text{m}$ and a length of $L_{\text{slot}} = 400\,\mu\text{m}$, equivalent to approximately half the guided wavelength at the targeted radiation frequency of 0.28 THz. The height of the structure ($h = 250\,\mu\text{m}$) was adapted such that the evaporated metal on the substrate simultaneously acts as a reflector. The measured radiation pattern for different elevation angles $\theta$ is shown in Fig. 5b, where the blue area represents the realized gain $G_{\text{R,TA,dBi}}(\theta) = 10\log_{10} G_{\text{R,TA}}(\theta)$ in the upper half-space region (*H*-plane). A maximum realized gain of 5.5 dBi was measured perpendicular to the substrate near 0.274 THz, in good agreement with the simulation (black dashed line). More details on the measurement setup and the characterization are given in Supplementary Section 6. To the best of our knowledge, our experiments represent the first demonstration of an additively manufactured 3D freeform mmW / THz antenna which does not require any manual assembly steps. Previous approaches have either been limited to the fabrication of bulky stand-alone horn antennas[31,32] or are based on aerosol-jet-printed conductive structures on the surface of non-planar dielectric substrates[52], often requiring additional mechanical assembly steps to yield a functional device[31,52].

## 4. Summary and outlook

We have introduced a novel concept for fabricating precisely defined mmW and THz structures, relying on *in-situ* printing of freeform polymer support structures that are selectively coated through highly directive metal deposition techniques. The resulting metal-coated freeform structures (MCFS) offer high surface quality in combination with conductivities comparable to bulk material values and do not require any manual assembly steps. The viability of the approach is shown in a series of proof-of-concept experiments. In a first set of experiments, we show THz interconnects (TIC) that bridge the gap between transmission lines located on different substrates and that offer 3 dB-bandwidths exceeding 0.33 THz. In a second set of experiments, we demonstrate that the vast design freedom offered by our fabrication technique can be leveraged for cost-effective THz probes (TP) that allow highly repeatable contacting over many probing cycles. Our TP offer 3 dB-bandwidths exceeding 0.19 THz and 6 dB-bandwidths far beyond the 0.33 THz range of our measurement system. A third set of experiments is finally dedicated to THz antennas (TA), which are suspended from the underlying high-index substrate for better radiation efficiency. At a frequency of 0.27 THz, a maximum realized gain of 5.5 dBi in the direction perpendicular to the substrate is measured. Our proof-of-principle experiments show the vast potential offered by the proposed fabrication technique, and the demonstrated structures already offer unique performance parameters such as record-high bandwidths for in-plane mmW chip-chip connections and for additively manufactured probes. Our concept offers unprecedented design freedom, does not rely on expensive high-precision assembly steps, and is widely applicable to a rich variety of use cases, thereby paving a path towards advanced mmW and THz systems in communications[35], sensing[36], or ultra-broadband signal processing[37].

## Methods

**CPW design and fabrication.** The substrates that are used as the base for the subsequent fabrication steps have a total size of 100 mm × 100 mm each and consist of a 635 µm-thick alumina ($Al_2O_3$) layer with a single-sided gold coating (Reinhardt Microtech GmbH, Ulm, Germany) and a thin titanium adhesion layer in between. The $Al_2O_3$ layer exhibits a relative permittivity of $\varepsilon' = 9.9$ and a loss tangent of $\tan(\delta) = 1 \times 10^{-4}$, both specified at a frequency of 1 MHz. The gold metallization has a thickness of 3 µm with a specified variance of ± 30%, and a root-mean-square (RMS) surface roughness of $R_{q,\mathrm{gold}} = (121...129)\,\mathrm{nm}$ was measured using a white-light interferometer, see Supplementary Section 3 for further information. For achieving a characteristic CPW impedance of 50 Ω, different combinations of the signal conductor width $w_S$, ground conductor width $w_G$ and signal-to-ground conductor distance $d_{SG}$ were chosen. All CPW designs facilitate contacting with GSG probes with pitches of 100 µm for compatibility with existing measurement equipment.

The transmission lines used in the experiments for the TIC and TA were structured into the gold layer using a wet etching process. To this end, a 1.5 µm-thick layer of positive-tone photoresist (AZ 1505, Microchemicals GmbH, Ulm, Germany) was first spin coated on the substrates and subsequently baked at a temperature of 95°C for five minutes. The resist was then structured using a direct-write laser lithography tool (DWL 66fs, Heidelberg Instruments Mikrotechnik GmbH, Heidelberg, Germany) and developed (AZ 400K 1:4, Microchemicals GmbH). To strip the metal in the uncoated regions, wet etching is performed with potassium iodide and titanium-tungsten etchants. Due to the significant variation of the metallization thickness across the substrate, over- and underetching of the metal structures was hardly avoidable. As a result, $w_S$ and $w_G$ were generally found to be decreased, while $d_{SG}$ was increased compared to the original design. The dimensions of the fabricated CPW are indicated in Table 1 and show fair agreement with the design values. From simulations, the characteristic impedance is expected to be in the range 58 Ω … 60 Ω. These limitations can be overcome by an optimized fabrication process, relying on gold layers with better thickness uniformity, which would further improve the impedance matching and hence the performance of our MCFS-based THz structures.

In contrast to this, the transmission lines used in the experiments for the TP were structured into the gold layer using a commercially available laser ablation system (ProtoLaser R4, LPKF Laser & Electronics SE, Garbsen, Germany), equipped with a 515 nm picosecond laser with a typical beam diameter of 15 µm. We used 1.5 ps-wide pulses with 24 µJ pulse energy at repetition rates between 150 kHz and 300 kHz for structuring the substrates. The dimensions of the fabricated CPW are again listed in Table 1 and show good agreement with the design values. The characteristic impedance is expected to be approximately 55 Ω.

**Table 1:** Overview of dimensions of manufactured CPW.

| Experiment | $L$ (mm) | $w_S$ (µm) | $w_G$ (µm) | $d_{SG}$ (µm) |
|---|---|---|---|---|
| TIC I | 4.0 | 45 | 65 | 30 |
| TIC II | 4.0 | 56 | 72 | 35 |
| TIC III | 4.0 | 30 | 150 | 30 |
| TP | 1.5 / 3.8 / 6.8 | 61 | 79 | 29 |
| TA | 1.1 | 22 | 89 | 16 |

**Sample preparation.** After structuring the CPW, a wafer saw was used for separation of the alumina substrates into individual chips. For easier handling, the individual chips were temporarily fixed to glass cover slides using UV glue (NOA 61, Norland Optical Adhesive Inc., Jamesburg, USA). For the experiments with the TIC, see Fig. 3, the ends of the 4 mm-long CPW feeds are approximately 230 µm away from the diced chip edges, keeping the length of the bridging TIC ($L_{\mathrm{TIC}} \approx 0.5\,\mathrm{mm}$) manageable when the CPW are placed facing each other. The gap between the two substrates was roughly 40 µm wide. The sample chip used in the TP experiments, see zoom-in in Supplementary Fig. S5, is roughly 4.4 mm wide and 8.8 mm long to provide sufficient area for pick-up with a custom vacuum tool (3.5 mm × 8 mm). The connecting CPW ($L_{\mathrm{conn}} = 3.8\,\mathrm{mm}$) between the two TP ends approximately 320 µm away from the diced chip edge on each side. This leaves enough room for a smooth routing of the signal and ground conductors, while ensuring that the tips of the TP protrude beyond the edges of the chip by a sufficient distance $d_{\mathrm{edge}}$ for better visibility via the top-view camera used in the experimental setup illustrated in Supplementary Fig. S5. In the areas between the ends of the transmission lines and the chip edge, the gold was removed from the alumina substrates for a more reliable height detection and better adhesion of the 3D-printed support structures, see next paragraph. Similarly, for the TA as shown in Fig. 5, the metal layer is removed from the surface of the substrate in a 250 µm-long region following each end of the transmission line. The remainder of the gold layer was maintained to act as a reflector for the suspended TA.

**Multi-photon lithography.** The 3D-printed support and shadowing structures, see Fig. 1b – d, were fabricated using an in-house-built lithography system with a 40× / 1.4 objective (PlanApochromat Oil DIC M27, Carl Zeiss Microscopy GmbH, Oberkochen, Germany), galvanometer-actuated mirrors, and a 780 nm femtosecond laser with a pulse width of 58 fs (C-Fiber 780 HP, Menlo Systems GmbH, Planegg, Germany). The samples were mounted in the lithography machine along with the associated glass cover slides, and the objective approaches the samples along the negative $z$-direction, see Figs. 3 – 5. For lithographic structuring, we use a liquid photoresist (VanCore B, Vanguard Automation GmbH, Karlsruhe, Germany), for which the dielectric response was independently measured, see Supplementary Section 1. All structures are printed to the substrates with the axis of the lithography beam perpendicular to the surface of the chips, i.e., along the $z$-direction, see Figs. 3 – 5. We use automated procedures for detecting the substrate surface with sub-100 nm precision. Where possible, the underlying metal layer was removed to avoid scattering of the laser beam due to the roughness of the gold layer. For lateral alignment, we relied on the camera-based vision system of our lithography setup, which leads to a lateral alignment precision that is also of the order of 100 nm. Due to the limited write field, corresponding approximately to a circle with a diameter of 400 µm, the support and shadowing structures have to be printed in separate steps. The support structures of the TIC and TA are further split into two and three parts, respectively, which are printed one after another. Proper alignment of the separately printed models is ensured by a defined translation of the underlying chip using a precision stage in the lithography system. To reduce the required fabrication time, all structures are printed with a standard layer-to-layer distance (slicing distance) of 100 nm only for the outermost shell, whereas the inside of the structures is filled with a coarser spacing of 600 nm. After printing, unexposed photoresist is removed in a two-step development process using propylene-glycol-methyl-ether-acetate (PGMEA) as a developer for 15 minutes, followed by rinsing in isopropyl alcohol (2-propanol).

**PMMA ink preparation and inkjet printing.** For protecting the planar THz structures during the global metal deposition, see Fig. 2c and e, we used an inkjet-printed poly(methyl methacrylate) (PMMA) layer, which was later removed by propylene-glycol-methyl-ether-acetate (PGMEA). Since PGMEA is also used in the development process of the 3D-printed structures, see Methods above, deterioration of the 3D-printed support structures during the PMMA lift-off is not to be expected. For the ink preparation, PMMA with an average molecular mass of $1.5 \times 10^4$ daltons (Sigma-Aldrich Inc., Saint Louis, USA) was dissolved in 1,3-dimethoxybenzene (≥98%, Sigma-Aldrich Inc.) to achieve a concentration of 80 g/L. We further added 5% hexylbenzene (97%, Sigma-Aldrich Inc.) to mitigate the coffee-ring effect[53]. Finally, the ink was filtered using polytetrafluoroethylene (PTFE) filters (pore size 200 nm) before being filled in the cartridges (FUJIFILM Dimatix Inc., Santa Clara, USA) of the inkjet printer (PiXDRO LP50, SÜSS MicroTec SE, Garching, Germany). The substrate and printhead temperatures were set to 50°C and 27°C, respectively. The build-in camera system of the printer was used to align the printing region with respect to the CPW and the shadowing structures on the chips. For all samples, the transmission lines are covered over their full length up to the flow-stop that is part of the 3D-printed shadowing structure, see Fig. 2c. To achieve sufficient coverage, a minimum of five layers were printed using a resolution of 550 dpi.

**Shadowing and metal deposition.** For applying the highly-directive metal coating, see Fig. 2d, an electron-beam physical vapor deposition (PVD) process is used (Univex 400, Leybold GmbH, Cologne, Germany) at a pressure of $8 \times 10^{-4}$ Pa. The sample chips are mounted with a large distance (> 60 cm) above the evaporation source, thereby achieving a small divergence of the vapor flow. The mounting plate carrying the samples can further be tilted in one direction and rotated around an axis that is normal to the mounting surface to improve the uniformity of the deposited layers. For the coating of the TIC and TP, the direction of evaporation is perpendicular to the substrates, i.e., along the negative z-direction in Figs. 2 – 4. For the TA in Fig. 5, the direction of deposition is set to an angle of 45° with respect to the negative z-direction in the (*y*,*z*)-plane, i.e., the deposition is done along the (0,1,-1)-direction of the coordinate system shown in Fig. 5a, which permits coating of the support structure also in the ramp-like sections which are partially perpendicular to the substrate. The deposited layer stack comprises a sequence of different metals: First, a 5 nm-thick adhesion film of titanium is applied, followed by the main layer consisting of 300 nm copper. Afterwards, a 5 nm titanium layer is deposited to act as a diffusion barrier between the copper and the final 10 nm-thick gold passivation layer. For the TIC and TA, the process is repeated twice for a total thickness of the metal coating of $h_\mathrm{m} = 0.64\,\mu\mathrm{m}$. For the TP, the initial titanium adhesion layer was replaced by a 30 nm aluminum coating, leading to a layer stack with a thickness of 345 nm. The process is repeated four times for a total metal thickness of $h_\mathrm{m} = 1.38\,\mu\mathrm{m}$ for the TP. This ensures that the thickness of the metal coating is always a multiple of the skin depth at the targeted operation frequency. For estimating the conductivity of our metal films, multiple metal strips with a length of 2 mm and width of 10 µm and 15 µm were fabricated on a silicon wafer with a 1 µm-thick silicon dioxide isolating layer on top (Microchemicals GmbH). We use identical evaporation parameters and repeat the process two times for achieving a metal layer stack with a thickness which was measured to be $h_\mathrm{m} \approx 590\,\mathrm{nm}$ using a profilometer (Tencor P-7, KLA Inc., Milpitas, USA). From four-wire measurements[39] using a precision source / measurement unit (B2902A, Keysight Technologies Inc., Santa Rosa, USA), we extract an effective sheet resistance of $R_\mathrm{S,MCFS} = (51.6 \pm 1.9)\,\mathrm{m}\Omega/\square$, corresponding to an effective conductivity of $\sigma_\mathrm{MCFS} = 1/(h_\mathrm{m} R_\mathrm{S,MCFS}) = (3.29 \pm 0.12) \times 10^7\,\mathrm{S/m}$. To mitigate offset errors as a result of the thermal electromotive force, measurements were carried out for both polarities of the driving current and by considering the average of the two associated voltage readings (reverse current method[39]).

**3D EM simulations.** The transmission and reflection characteristics of the TIC, see Fig. 3c and d, the TP, see Fig. 4b, as well as the far-field characteristics of the TA, see Fig. 5b, are simulated using a commercially available numerical time-domain solver (CST Microwave Studio, Dassault Systèmes SE, Vélizy-Villacoublay, France). The alumina substrate is modeled using the manufacturer specifications of the relative permittivity and loss tangent, see paragraph on "CPW design and fabrication" above. The CPW are simulated using the "lossy metal" option of CST Microwave Studio, which models the penetration of electromagnetic fields inside a very good but still imperfect conductor by means of an internal one-dimensional surface impedance model[54]. This allows to take into account the skin effect without further mesh refinement. We also include the measured RMS surface roughness $R_\mathrm{q,gold}$ of the gold layer of the respective assembly, see above. This value is used as an input for the so-called gradient model provided within CST Microwave Studio, which accounts for the increased total loss and inner inductance effects that occur as a consequence of the rough surface[54]. The material characteristics of the 3D-printed support structures are modeled using the results of the dielectric response measurement of the underlying photoresist, see Supplementary Section 1. To this end, the measured values for the permittivity $\varepsilon'$ and loss tangent $\tan(\delta)$ in the frequency range from 0.220 THz to 0.325 THz were imported into CST Microwave Studio and fitted using a dispersion model of first order. Another crucial aspect is the correct representation of the comparatively small isolation trenches inside the 3D-printed support structures, separating the deposited metal layers by a gap of the order of a view micrometers. In the simulation of transmission lines and THz devices with dimensions up to the mm-range, a detailed representation of the field within these gaps and metal layers would lead to an unrealistic number of mesh cells that cannot be handled with the available computing resources. We therefore use a locally refined mesh inside the gaps in combination with the "thin panel" option of CST Microwave Studio, which relates the tangential electric and magnetic fields on the surface of the extended metal strips with the help of surface impedances[54]. Furthermore, the TA is simulated with open boundaries in combination with a so-called multilayer background to emulate an infinitely extended substrate. Using this approach, any reflections at the lateral edges of the substrate can be avoided, albeit the restricted simulation volume.

**Micrograph image acquisition.** The micrographs in Fig. 1b – d and in Fig. 3b are created using a digital microscope (VHX-7000, Keyence Ltd., Osaka, Japan) with a tiltable and motorized imaging head comprising a CMOS camera (VHX-7100, Keyence Ltd.). A magnification of 300× and 150× was used for the micrographs in Fig. 1b, c and Fig. 1d, respectively, and the micrograph in Fig. 3b was taken at a magnification of 80×. In all cases, the microscope was set to acquire large image stacks (~50 – 100 images) with different focus distances, which were then combined into focus-stacked images using a commercially available software (Helicon Focus, Helicon Soft Ltd., Kharkiv, Ukraine).

**Funding.** This work was supported by the Deutsche Forschungsgemeinschaft (DFG, German Research Foundation) via the Excellence Cluster 3D Matter Made to Order (EXC-2082/1 – 390761711), via the DFG Collaborative Research Center HyPERION (SFB 1527), and via the projects PACE (# 403188360) and GOSPEL (# 403187440) within the Priority Programme "Electronic-Photonic Integrated Systems for Ultrafast Signal Processing" (SPP 2111), by the ERC Consolidator Grant TeraSHAPE (# 773248), by the EU Horizon 2020 project TeraSlice (# 863322), by the European Innovation Council (EIC) transition project HDLN (#101113260), by the BMBF project Open6GHub (# 16KISK010), by the Alfried Krupp von Bohlen und Halbach Foundation, and by the Karlsruhe School of Optics & Photonics (KSOP).

**Author contributions.** The approaches and experiments were conceived by P.M., A.K., J.H., T.H., and C.K.. P.M. developed the lithography techniques and processes required for fabrication of the 3D-printed metal-coated freeform structures. The THz devices were designed by P.M. with the support of A.K., C.B., and T.H.. A.Q. supported the manufacturing of the transmission lines using the laser ablation system. A.Q. and M.K. supported the measurement of the dielectric response of the photoresist. The THz assemblies were fabricated by P.M., with support of Q.Z. for the inkjet-printing of the PMMA layers. P.M. performed the experimental characterization of the THz assemblies and analyzed the data, with support of A.K., J.H., and C.K.. All authors analyzed the results and discussed the data. P.M., W.F., and C.K. wrote the manuscript. The project was supervised by S.R., U.L., W.F., T.Z., and C.K..

**Disclosures.** C.K. is a co-founder and shareholder of Vanguard Photonics GmbH, Vanguard Automation GmbH, and Keystone Photonics GmbH, companies engaged in exploiting 3D nano-printing in the field of photonic integration, optical assembly, and microfabrication. P.M., A.K., A.Q., M.K., T.H., T.Z., and C.K. are co-inventors of patents owned by Karlsruhe Institute of Technology in the technical field of the publication. The other authors J.H., Q.Z., C.B., S.R., U.L. and W.F. declare no conflict of interest.

**Data availability.** The data that support the findings of this study may be obtained from the authors upon reasonable request.

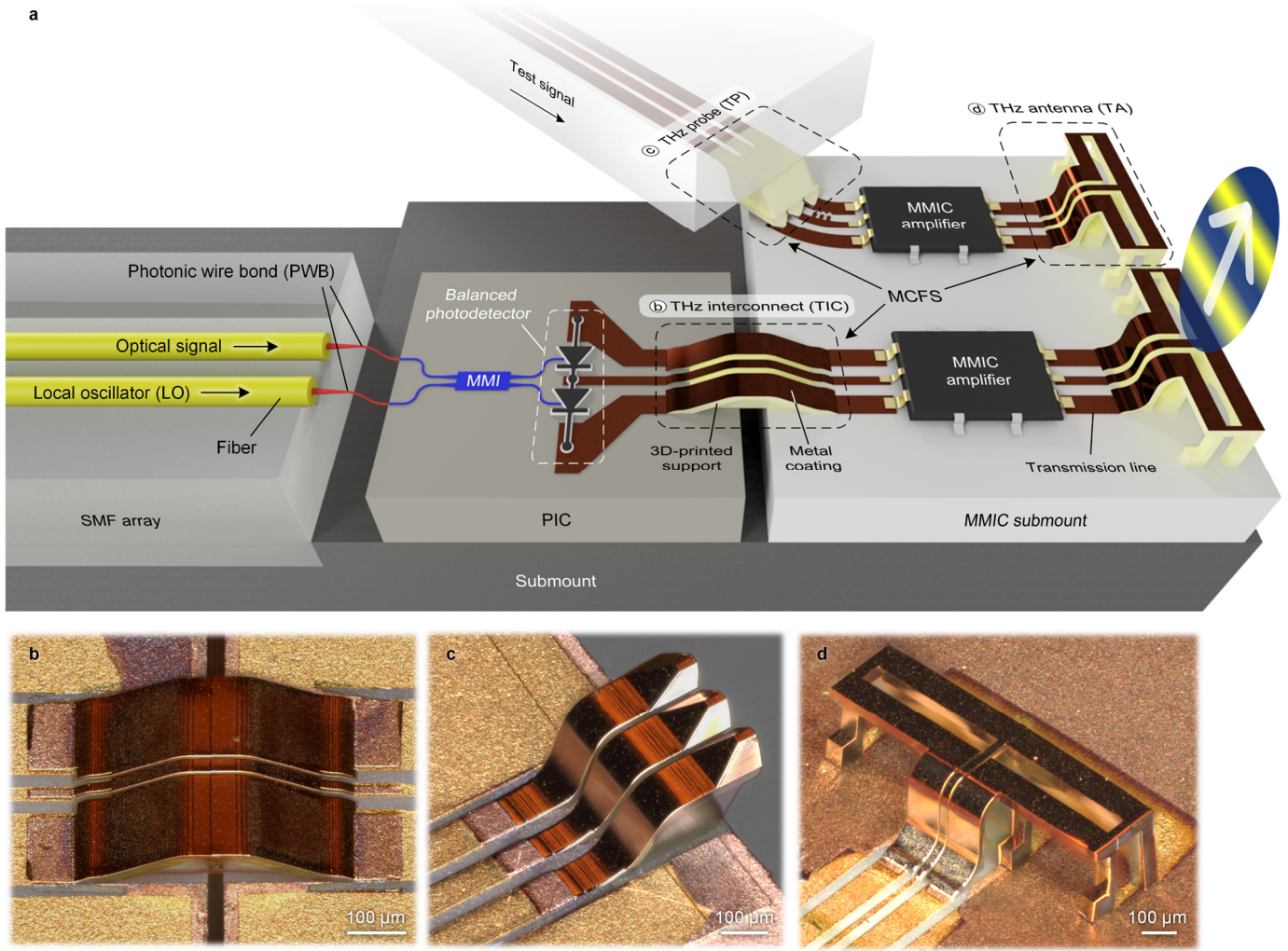

**Fig. 1 | Concept of a THz system exploiting 3D-printed functional elements based on metal-coated freeform structures (MCFS).**
**a,** The system combines an optoelectronic signal source, relying on a balanced pair of high-speed photodetectors that are part of a photonic integrated circuit (PIC), with a subsequent THz amplifier, which is based on a monolithic microwave integrated circuit (MMIC), and a THz antenna, which is suspended from the surface of the underlying substrate for efficient emission to the surface-normal direction. The THz chip-chip interconnect between the PIC and MMIC as well as the suspended THz antenna are implemented as metal-coated freeform structures (MCFS). Each MCFS consists of a 3D-printed polymeric support that is locally coated with metal layers offering high bulk conductivity along with good surface quality. The same concept can be used to implement THz probes with unprecedented shape fidelity that lend themselves to testing of mmW and THz integrated circuits, see upper part of the figure. The polymeric support structures of the various MCFS are fabricated using high-resolution multi-photon lithography and can thus be efficiently combined with 3D-printed optical connections such as photonic wire bonds (PWB).
**b,** Micrograph of a THz interconnect (TIC) bridging the gap between coplanar waveguide transmission lines on different THz substrates.
**c,** Micrograph of a THz probe (TP) that is turned upside down for better visibility. **d,** Micrograph of a THz antenna (TA) that is suspended from the underlying high-index substrate for increased radiation efficiency.

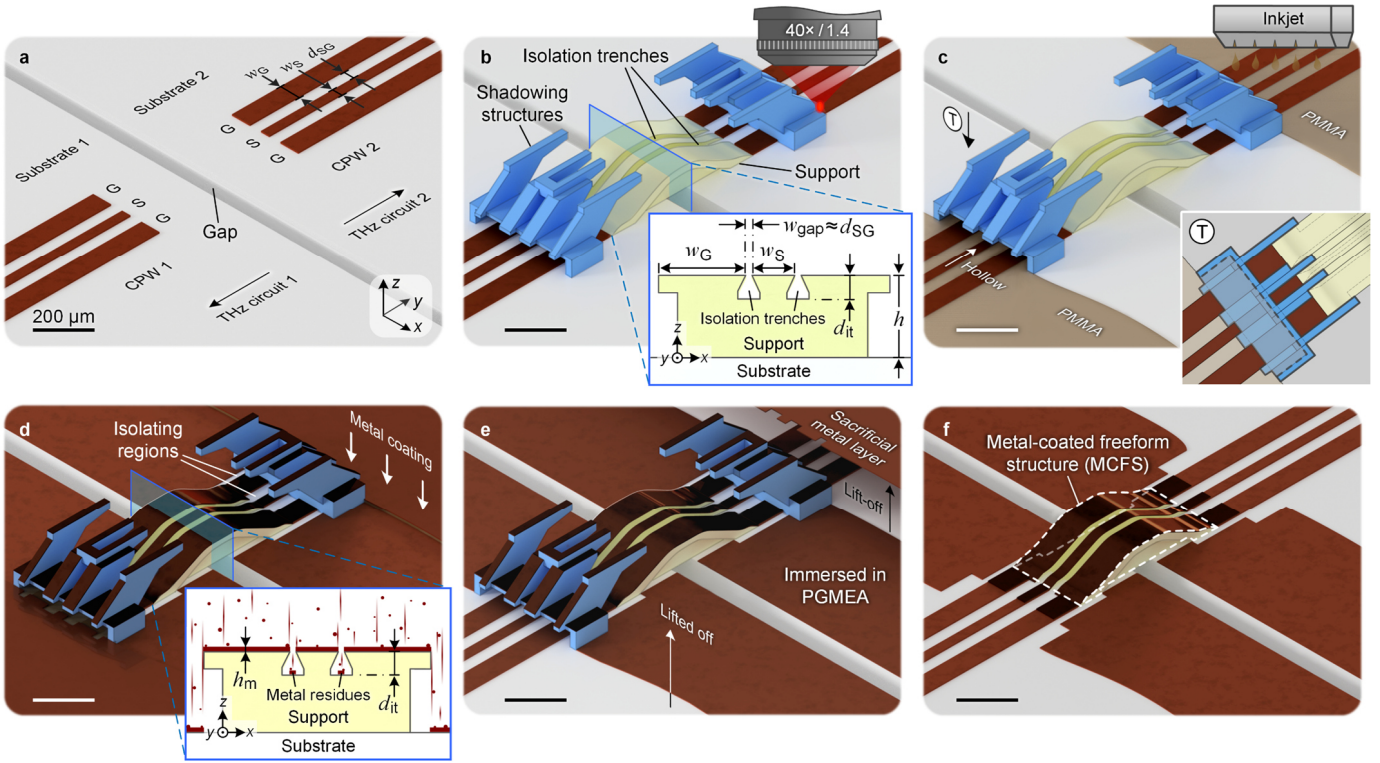

**Fig. 2 | Fabrication of metal-coated freeform structures (MCFS).** The process steps are shown exemplarily for the fabrication of a THz interconnect (TIC) as depicted in Fig. 1b. All scale bars correspond to 200 µm. **a,** Two substrates with prefabricated planar THz structures such as coplanar waveguides (CPW) in ground-signal-ground (GSG) configuration are coarsely placed to face each other with a gap in-between. **b,** Freeform support and shadowing structures are 3D-printed using *in-situ* multi-photon laser lithography. The structures are designed to support a metal layer which connects smoothly to the planar CPW on each substrate and comprise isolation trenches with undercut sidewalls, see Inset. These trenches are designed to separate the deposited GSG metal strips, see Subfigure d. At the transitions between the planar substrates and the freeform polymeric support, additional 3D-printed shadowing structures are used to prevent short circuits. These shadowing structures take the form of multiple arm-like roofs, which locally prevent metal deposition, thereby defining isolating regions that separate the metal strips of the GSG transmission line and that partially overlap with the isolation trenches of the adjacent 3D support structure, see Subfigures c and d. The shadowing structures are shaded in blue. **c,** For protecting the planar THz structures during global metal deposition, the associated areas of the substrates are temporarily covered by a PMMA film using inkjet printing. Inset ⓣ provides a top-view on one side of the support structure with the corresponding shadowing structure mentioned above. The shadowing structure is built upon a box-like hollow base, which acts as a flow-stop preventing inkjet-printed PMMA from wetting the entire support structure. **d,** A highly directed physical vapor deposition (PVD) process is used to deposit metal along a surface-normal direction, forming highly conductive layers on all surfaces with direct line of sight to the evaporation source. **e,** After metal deposition, the sacrificial PMMA layer is dissolved, thereby lifting off the unwanted metal areas. **f,** As a last step, the 3D-printed shadowing structures can be removed mechanically, leaving the final MCFS.

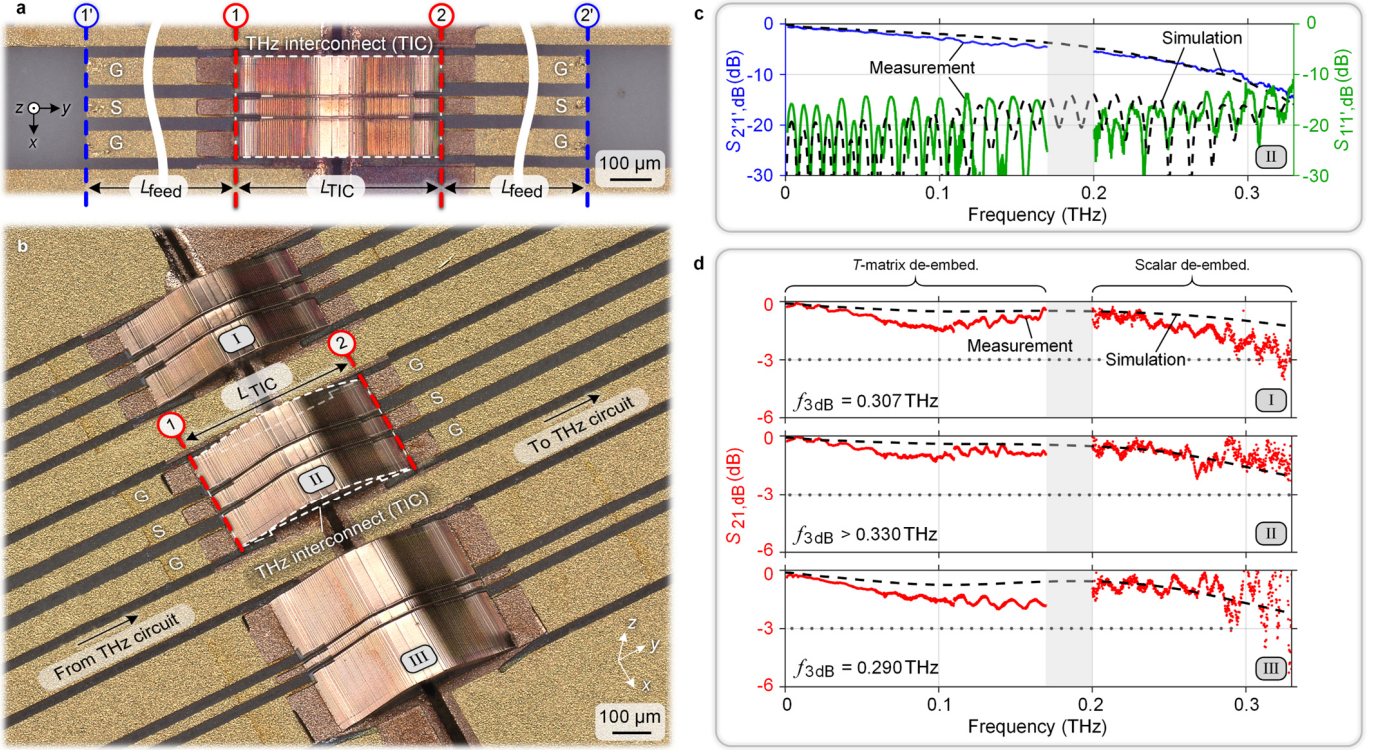

**Fig. 3 | Characterization of the THz interconnects (TIC). a,** Top view of a manufactured TIC, seamlessly connecting the GSG strips of two CPW on two alumina substrates. The CPW endings are separated by $L_{\mathrm{TIC}} \approx 0.5\,\mathrm{mm}$, and the length of the feed lines amounts to $L_{\mathrm{feed}} = 4\,\mathrm{mm}$. For characterization of the TIC, we first measure the compound *S*-parameters of the TIC and the associated feed lines, which are contacted using dedicated probes in Plane $1'$ and Plane $2'$ (blue lines), see Subfigure c. We then apply a de-embedding procedure to obtain the *S*-parameters of the TIC only, associated with Plane $1$ and Plane $2$ (red lines), see Subfigure d. **b,** Micrograph of three manufactured TIC, labeled I, II and III. **c,** Compound *S*-parameters of TIC II including the associated feed lines. The blue solid curve refers to the measured transmission $S_{2'1',\mathrm{dB}} = 10\log_{10}(|\underline{S}_{2'1'}|^2)$, and the green solid curve to the measured reflection $S_{1'1',\mathrm{dB}} = 10\log_{10}(|\underline{S}_{1'1'}|^2)$. The measurements show excellent agreement with a simulation of the associated structure (dashed black lines). **d,** To remove the influence of the feed lines, we measure the *S*-parameters of a reference CPW on the same substrate and apply de-embedding procedures. The associated transmission characteristics $S_{21,\mathrm{dB}} = 10\log_{10}(|\underline{S}_{21}|^2)$ (red curves) are shown for TIC I (upper), II (center) and III (lower), revealing 3 dB-bandwidths of 0.307 THz, >0.330 THz and 0.290 THz, respectively. Measurements and simulations (dashed black lines) agree again well. The measured *S*-parameters are subject to a gap between 0.170 THz and 0.200 THz due to lack of adequate signal sources. For frequencies up to 0.170 THz, a transfer-matrix (*T*-matrix) approach was used for de-embedding, whereas for frequencies beyond 0.200 THz, the de-embedded *S*-parameters were estimated using a more robust scalar correction that is less prone to error multiplication, see Supplementary Section 4 for details.

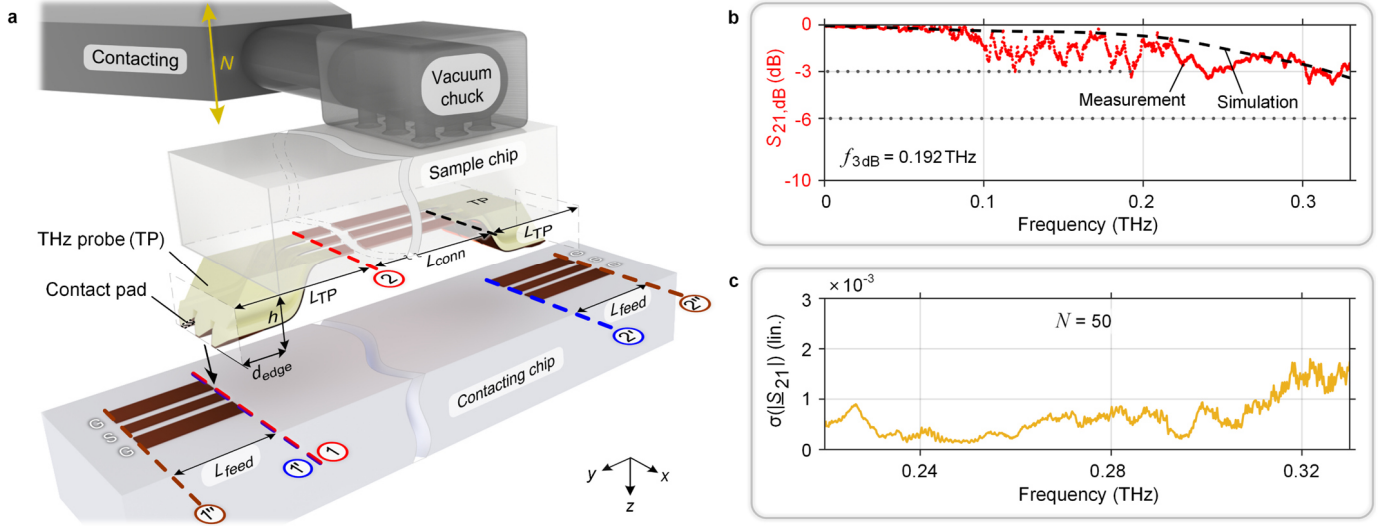

**Fig. 4 | Characterization of the THz probes (TP). a,** Identical twins of TP ($L_{\mathrm{TP}} = 415\,\mu\mathrm{m}$, $h = 180\,\mu\mathrm{m}$) are fabricated on a sample chip at opposite ends of a CPW ($L_{\mathrm{conn}} = 3.8\,\mathrm{mm}$). The three tips (pitch 100 µm) are design for a contact pad size of $20 \times 40\,\mu\mathrm{m}^2$ or more and protrude beyond the edges of the sample chip by a distance $d_{\mathrm{edge}}$. A second chip with two short CPW feeds ($L_{\mathrm{feed}} = 1.5\,\mathrm{mm}$) was used for contacting (contacting chip), where the distance between the endings of the CPW are slightly smaller than $L_{\mathrm{TP}} + L_{\mathrm{conn}} + L_{\mathrm{TP}}$. A vacuum chuck attached to a manual actuator is used to move the sample chip in the desired position above the contacting chip. With the sample chip in contact, $S$-parameters between Plane $1''$ and Plane $2''$ (brown lines) are measured. Using dedicated de-embedding procedures, the reference planes are first moved to Plane $1'$ and Plane $2'$ (blue lines) and then to Plane $1$ and Plane $2$ (red lines), thereby extracting the characteristics of a single TP. **b,** Estimated transmission characteristics $S_{21,\mathrm{dB}} = 10\log_{10}(|\underline{S}_{21}|^2)$ of a single TP. The 3 dB-bandwidth amounts to 0.192 THz, and the 6 dB-bandwidth is far beyond the 0.330 THz range of our measurement system. Measurement and simulation (dashed black line) are again in reasonable agreement. **c,** To further demonstrate the robustness of our probes and to verify the repeatability of the connections, we repeat our measurement $N = 50$ times, detaching and re-establishing the contact between the sample and the contacting chip after each repetition. From the resulting data, we extract the standard deviation of the magnitude of the complex-valued amplitude transmission factor $\underline{S}_{21}$ as a function of frequency. We find a remarkably low maximum standard deviation of $\sigma(|\underline{S}_{21}|)_{\max} \approx 2 \times 10^{-3}$ near 0.32 THz, corresponding to –55 dB.

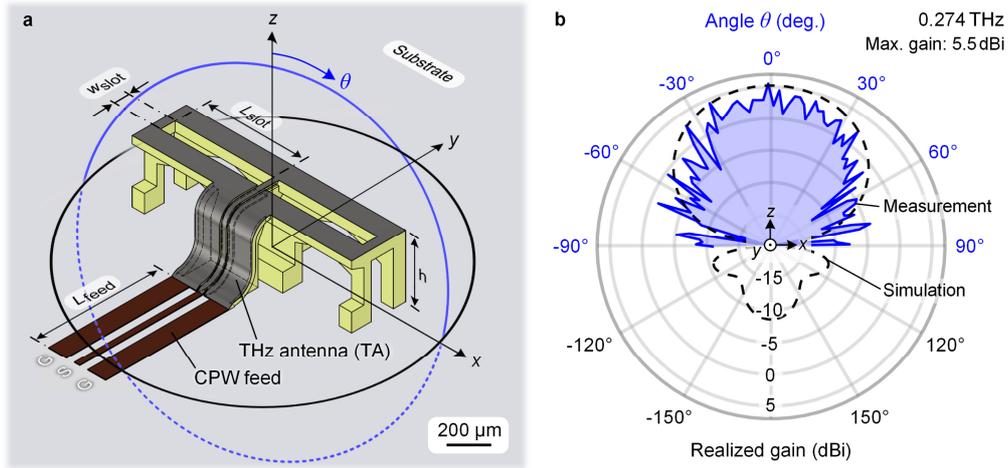

**Fig. 5 | Characterization of the THz antennas (TA). a,** Schematic of the manufactured TA operating at a frequency centered around 0.28 THz. Connecting to the CPW feed ( $L_{\text{feed}} = 1.1\,\text{mm}$ ), the 3D-printed support structure provides a mechanical base for guiding the signal and ground conductors of the CPW away from the substrate to form an elevated slot antenna for improved radiation to the top. The antenna comprises two slots, each having a width of $w_{\text{slot}} = 60\,\mu\text{m}$ and a length of $L_{\text{slot}} = 400\,\mu\text{m}$, equivalent to approximately half the guided wavelength at the targeted radiation frequency of 0.28 THz. The height of the support structure $(h = 250\,\mu\text{m})$ was adapted such that the evaporated metal on the substrate simultaneously acts as a reflector. **b,** Measured radiation pattern for different elevation angles $\theta$, where the blue area represents the realized gain $G_{\text{R, TA,dBi}}(\theta) = 10\log_{10} G_{\text{R,TA}}(\theta)$ in the upper half-space region (*H*-plane). A maximum realized gain of 5.5 dBi was measured perpendicular to the substrate near 0.274 THz, in good agreement with the simulation (black dashed line).